\documentclass[aps, twocolumn, superscriptaddress]{revtex4-2}

\usepackage{graphicx}
\usepackage{dcolumn}
\usepackage{bm}
\usepackage{hyperref}
\usepackage[utf8]{inputenc}
\usepackage[T1]{fontenc}
\usepackage{mathptmx}
\usepackage{etoolbox}
\usepackage{amssymb}
\usepackage{amsmath}
\usepackage{amsthm}
\usepackage{colortbl}
\usepackage[table,xcdraw]{xcolor}
\usepackage{amsmath}

\begin{document}

\setlength{\parindent}{0.5cm}



\title{On forced swarmalators that move in higher-dimensional spaces}

\author{Md Sayeed Anwar}
\email{sayeedanwar447@gmail.com}
\affiliation{Physics and Applied Mathematics Unit, Indian Statistical Institute, 203 B. T. Road, Kolkata 700108, India}

\author{Dibakar Ghosh}
\email{diba.ghosh@gmail.com (corresponding author)}
\affiliation{Physics and Applied Mathematics Unit, Indian Statistical Institute, 203 B. T. Road, Kolkata 700108, India}

\author{Kevin O'Keeffe}
\email{kevin.p.okeeffe@gmail.com}
\affiliation{Starling Research Institute, Seattle, USA} 


\begin{abstract}
We study the collective dynamics of swarmalators subjected to periodic (sinusoidal) forcing. Although previous research focused on the simplified case of motion in a one-dimensional (1D) periodic domain, we extend this analysis to the more realistic scenario of motion in two and three spatial dimensions with periodic boundary conditions. In doing so, we identify analogues of the 1D states and characterize their dynamics and stability boundaries analytically. Additionally, we investigate the forced swarmalators model with power-law interaction kernels, finding that the analytically tractable model with periodic boundary conditions can reproduce the observed dynamic behaviors of this more complex model.  \\


\noindent
\end{abstract}

\maketitle

\section{Introduction}

Swarmalators are mobile oscillators that serve as models for a great diversity of systems that simultaneously self-assemble in space and self-synchronize internal phase variables in time. Examples include sperm \cite{yang2008cooperation,creppy2016symmetry}, vinegar eels \cite{quillen2021metachronal,quillen2022fluid}, Japanese tree frogs \cite{aihara2014spatio}, magnetic domain walls \cite{hrabec2018velocity}, embryonic cells \cite{tsiairis2016self}, active spheres \cite{riedl2023synchronization,romano2024entropy}, active spinners \cite{sungar2024synchronization}, Janus particles \cite{yan2012linking}, Quincke rollers \cite{zhang2020reconfigurable}, and robotic swarms \cite{barcis2019robots,barcis2020sandsbots,monaco2020cognitive}. 


The study of swarmalators began a decade or so ago \cite{tanaka2007general,o2017oscillators} and since then has received growing interest. The first studies considered the simplest possible models, such as those with interactions of uniform strength. Several collective states were found here whose analysis is an open problem \cite{o2024solvable}. Later studies modified this baseline model in various ways. The effects of more realistic coupling structures \cite{blum2024swarmalators,sar2022swarmalators,lizarraga2023synchronization}, random pinning \cite{sar2023pinning,sar2023swarmalators,sar2024solvable}, thermal noise \cite{hong2023swarmalators}, distributed natural frequencies \cite{yoon2022sync}, and higher-order interactions \cite{anwar2024collective} have been studied. 

This paper is about swarmalators subject to external forcing, an effect which occurs in many real-world systems. Magnetic colloids are a prime example. If you shine the colloids with external fields, the units' dipole begin to oscillate and self-synchronize. This couple to the unit's movement, and in that sense allows for a kind of sync-mediated control. This control has marvelous use-cases. It has been used to degrade pollutants \cite{ursobreaking,dai2021solution,vikrant2021metal,tesavr2022autonomous}, repair the cracks in electrical circuits \cite{li2015self}, and to break down blood clots \cite{cheng2014acceleration,manamanchaiyaporn2021molecular}. 

A theoretical understanding of forced swarmalators like the above systems is lacking. Lizarraga et al. \cite{lizarraga2023synchronization} took the first step with a numerical study of the two-dimensional (2D) swarmalator model  modified with sinusoidal forcing. They found diverse dynamics, but, due to the complexity of the 2D model \footnote{see the Introduction in \cite{o2024solvable} for a discussion about the analytic difficulties of the 2D model}, were unable to analyze them. In an effort to make some analytic progress on this class of problems, we recently studied the simpler 1D swarmalator model which confines the swarmalators' motion to a 1D periodic domain \cite{anwar2024forced}. We found rich collective dynamics which, in this simpler 1D setting, were tractable. 

This paper continues this effort. We scale up from one to two and then to three spatial dimensions, and study a model that imitates the behavior of the original swarmalator while remaining tractable. Specifically, we enforce periodic boundary conditions and omit excluded volume interactions. These simplifications let us derive stability curves and bifurcations of several of the model's collective states analytically. 

To our knowledge, these are the first analytical results about swarmalator with forcing in two and three spatial dimensions, and in that sense contribute to the field.

\section{Mathematical Models}
Below, we enumerate the different models and the phenomena they generate.

\begin{figure*}[t!]
	\centering
	\includegraphics[width=\linewidth]{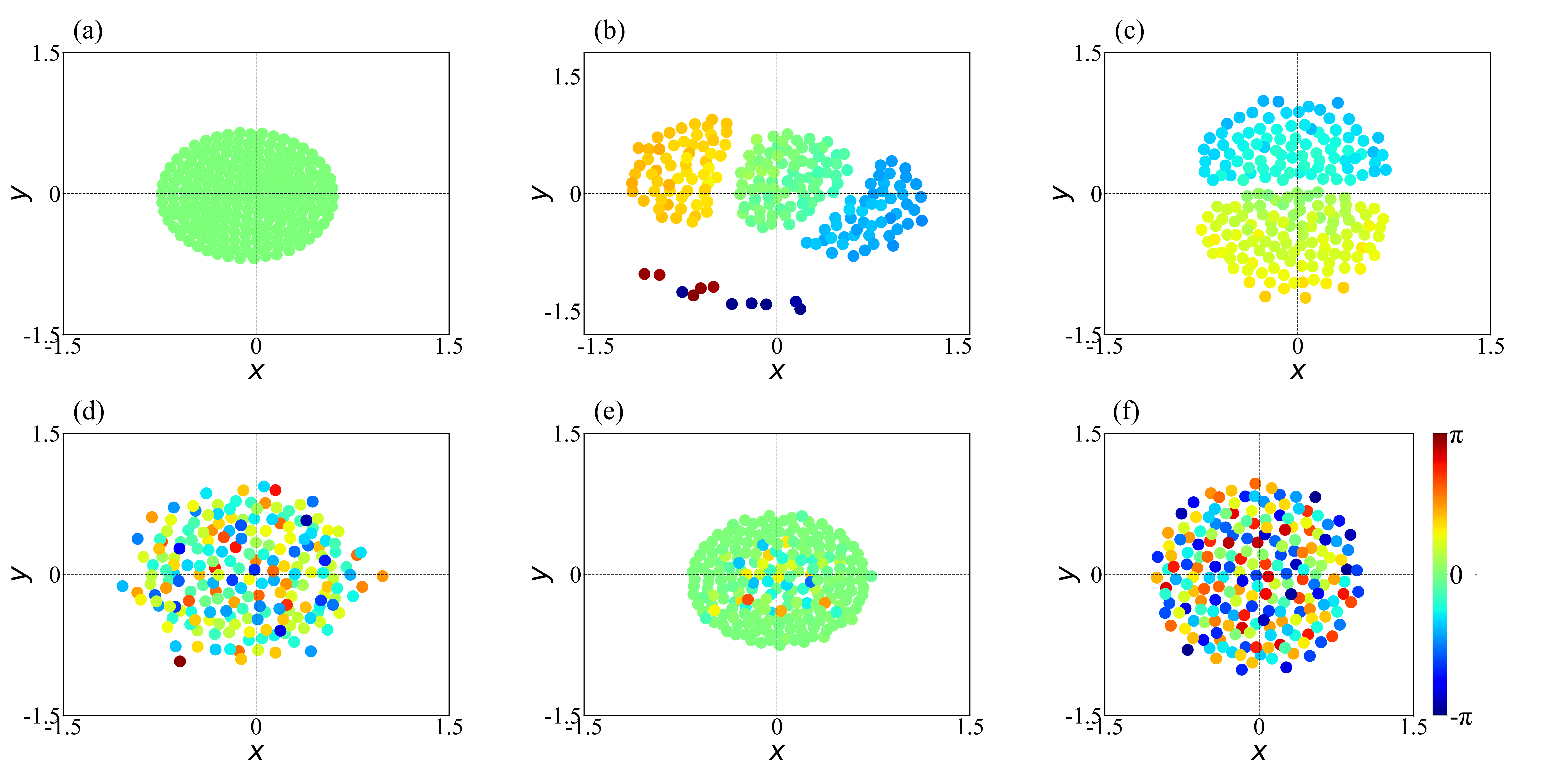}
	\caption{Collective states of the 2D swarmalator model. All the states are obtained for spatial coupling $J=1$. (a) Pinned state for $(K, F)=(-3,8.1)$, (b) multiple cluster state for $(K, F)=(-0.2,0.1)$, (c) two cluster state for $(K, F)=(-0.2, 0.15)$, (d) state with two coherent phase group for $(K, F)=(-3,2.5)$, (e) phase locked state for $(K, F)=(-3,6.5)$, and (f) unsteady incoherence state for $(K,F)=(-3,0.5)$. }
	\label{2d_model_states}
\end{figure*}

\textit{2D swarmalator model}. 
\begin{equation}
\begin{array}{l}
\dot{\mathbf{x}}_i = \nu + \dfrac{1}{N} \sum\limits_{ j \neq i}^N \Bigg[ \dfrac{\mathbf{x}_j - \mathbf{x}_i}{|\mathbf{x}_j - \mathbf{x}_i|^p} \Big( 1 + J \cos(\theta_j - \theta_i)  \Big) -   \dfrac{\mathbf{x}_j - \mathbf{x}_i}{ | \mathbf{x}_j - \mathbf{x}_i|^q}\Bigg],\\ \\
\dot{\theta_i} = \omega + \dfrac{K}{N} \sum\limits_{j \neq i}^N \dfrac{ \sin(\theta_j - \theta_i)}{ |\mathbf{x}_j - \mathbf{x}_i|^r } + F \sin(\Omega t - \theta),\; i=1,2,\dots,N. 
\end{array} \label{2d_model_eq}
\end{equation}
Here $\mathbf{x}=(x,y) \in \mathbb{R}^2$ is the position and $\theta \in S^1$ is the phase. $(\nu,\omega)$ are the free velocity and intrinsic frequency. The spatial dynamics model aggregation which depends on phase similarity. The power law kernels model long range attraction and short range repulsion (we require $q>p$ for this) and the $J \cos() $ term couples the space and phase dynamics. The phase dynamics are a generalized Kuramoto model with a sinusoidal forcing term added in. The parameters $J, K$ are the associated coupling strengths, and $F,\Omega$ are the strength and frequency of the forcing. For our study we consider a particular instance of this model with $p=0$, $q=2$ and $r=1$, which leaves the system with linear attraction kernel \cite{o2017oscillators}. Figure \ref{2d_model_states} shows some of its collective states.

\begin{figure*}[t!]
	\centering
	\includegraphics[width=\linewidth]{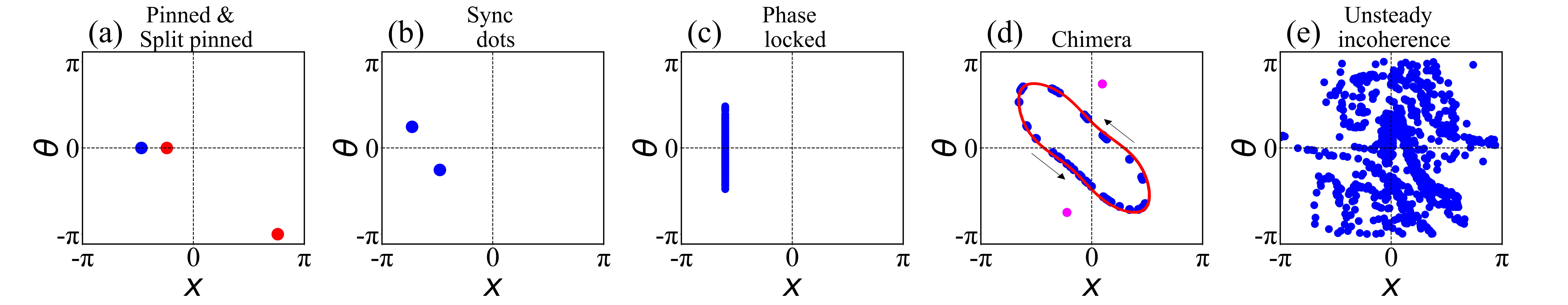}
	\caption{Collective states of the 1D swarmalator model at $J=1$ (reproduced from Ref.~\cite{anwar2024forced}). (a) Pinned (blue dot) and split pinned (red dots) state for $(K, F)=(1,0.5)$, (b) sync dots for $(K, F)=(-2,1)$, (c) phase locked state for $(K, F)=(-2, 1.8)$, (d) chimera state for $(K, F)=(-2,0.7)$, and (e) unsteady coherence state for $(K, F)=(-2,0.2)$.}
	\label{1d_model_states}
\end{figure*}

\textit{1D swarmalator model}. This considers swarmalators which are confined to move on a 1D periodic domain
\begin{equation}
\begin{array}{l}
    \dot{x}_i =  \nu + \dfrac{J}{N} \sum\limits_{j=1}^N \sin(x_j - x_i) \cos(\theta_j - \theta_i),  \\
    \dot{\theta}_i = \omega + \dfrac{K}{N} \sum\limits_{j=1}^N \sin(\theta_j - \theta_i) \cos(x_j - x_i)  + F \sin(\Omega t - \theta_i).
\end{array} \label{1d_model_eq}
\end{equation}
Here, $x_i, \theta_i \in S^1$ denote the position and phase of the $i$-th swarmalator ($S^1$ is the unit circle). The parameters $J, K$ are as usual the coupling parameters associated with space and phase interactions and $(\nu, \omega)$ are natural frequencies. These equations model aggregation in space which depends on phase similarity, and synchronization which depends on distance. The simplicity of the model makes it the one of only models of forced swarmalators that is tractable \cite{o2022collective, anwar2024forced, o2024global, o2024stability}. It can be derived from the 2D swarmalator model. You can think of it as this captures the rotational piece of the motion \cite{o2022collective}. Figure \ref{1d_model_states} shows it collective states. The details analysis of this model is presented in Ref.~\cite{anwar2024forced}.

\textit{2D swarmalator model with periodic boundary conditions}. 
\begin{equation}
\begin{array}{l}
	\dot{x}_i = u + \dfrac{J}{N} \sum\limits_{j} \sin(x_j - x_i) \cos(\theta_j - \theta_i), \\
	\dot{y}_i = \nu + \dfrac{J}{N} \sum\limits_{j} \sin(y_j - y_i)\cos(\theta_j - \theta_i), \\
	\dot{\theta}_i = \omega + F\sin(\Omega t - \theta_i)+\dfrac{K}{N} \sum\limits_{j} \sin(\theta_j - \theta_i) \Big[\cos(x_j - x_i) \\
	   \;\;\;\;\;\;\;\;\;\;\;\;\;\;\;\;\;\;\;\;\;\;\;\;\;\;\;\;\;\;\;\;\;\;\;\; +\cos(y_j - y_i) \Big].  
\end{array} \label{2d_torus_model_eq}
\end{equation}
This is a generalization of the 1D model above, where now the swarmalators run in the plane with periodic boundary conditions $(x_{i},y_{i},\theta_{i} \in S^{1})$. The space equations capture spatial attraction that depend on phase similarity. The phase equation illustrates synchronization that depends on distance. You can think of the $\cos(x_j - x_i) + \cos(y_j - y_i)$ are a distance metric \cite{o2024solvable}. 

\begin{figure*}[t!]
	\centering
	\includegraphics[width=\linewidth]{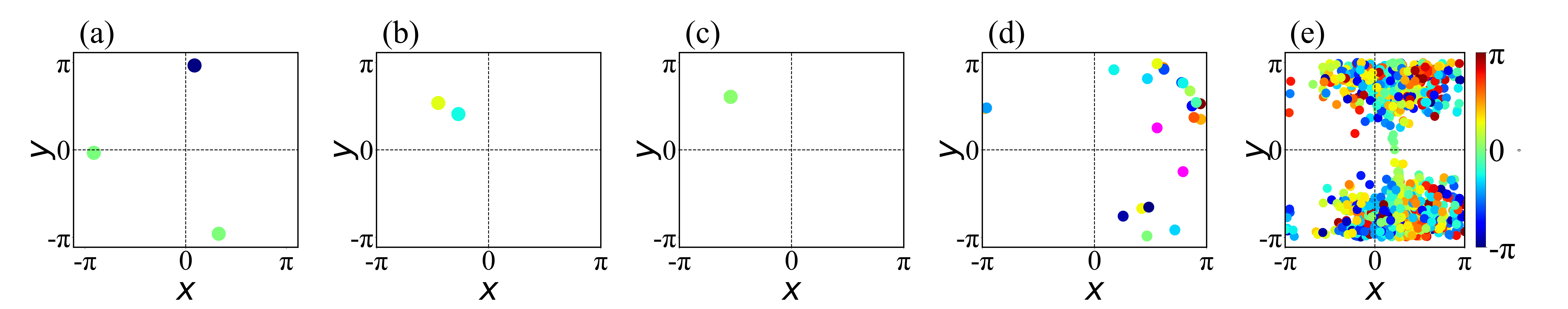}
	\caption{Collective states of the 2D swarmalator model with periodic boundary conditions. (a) Pinned (right green dot) and split pinned (blue and left green dots) state for $(K, F)=(1,0.3)$, (b) sync dots for $(K, F)=(-2,2.5)$, (c) phase locked state for $(K, F)=(-2, 3.5)$, (d) chimera state for $(K, F)=(-2,2.1)$, pink dots are the sync populations of the chimera state and (e) unsteady coherence state for $(K, F)=(-2,0.3)$. The states are obtained for spatial coupling $J=1$.}
	\label{2d_torus_states}
\end{figure*}

Notice the model does not include hard shell repulsion, like the 2D swarmalator model. This convenient feature makes it one of the few -- to our knowledge the only -- model of swarmalators that is tractable. When the forcing is turned off, its  phase diagram and order parameters may be derived exactly \cite{o2024solvable}. Generalizing these results when the forcing is turned on is the subject of this paper. Figure \ref{2d_torus_states} shows its collective states.

\textit{3D swarmalator model with periodic boundary conditions}. 
\begin{equation}
\begin{array}{l}
	\dot{x}_i = u+ \dfrac{J}{N} \sum\limits_{j} \sin(x_j - x_i) \cos(\theta_j - \theta_i), \\
	\dot{y}_i = \nu + \dfrac{J}{N} \sum\limits_{j} \sin(y_j - y_i)\cos(\theta_j - \theta_i) , \\
    \dot{z}_i = w + \dfrac{J}{N} \sum\limits_{j} \sin(z_j - z_i)\cos(\theta_j - \theta_i) , \\
	\dot{\theta}_i = \omega + F\sin(\Omega t - \theta_i)+\dfrac{K}{N} \sum\limits_{j} \sin(\theta_j - \theta_i) \Big[\cos(x_j - x_i)  \\
	\;\;\;\;\;\;\;\;\;\;\;\;\;\;\;\;\;\;\;\;\;\;\;\;\;\;\; +\cos(y_j - y_i) +\cos(z_j - z_i) \Big].  
\end{array} \label{3d_torus_model_eq}
\end{equation}
This is a generalization of the 2D model with periodic boundary conditions to three-dimensional (3D) space, and thus does not include the hard shell repulsion among the swarmalators. Here $x_{i},y_{i},z_{i} \in S^{1}$ denotes the position and $\theta_{i}\in S^{1}$ is the phase of $i$-th swarmalator. We can think of the $\cos(x_j - x_i) + \cos(y_j - y_i)+ \cos(z_j - z_i)$ as the distance metric. This model exhibits similar dynamics as the 2D model with periodic boundary conditions.  
\begin{table*}[t!]
\centering
\rowcolors{1}{gray!25}{white}
\begin{tabular}{|l|c|c|c|c|} 
\hline 
\rowcolor{gray!50} 
\textbf{State} & \textbf{2D Model} & \textbf{1D Model} & \textbf{2D Model Periodic BC} & \textbf{3D Model Periodic BC} \\ 
\hline 
Pinned & \cellcolor{green!30}\checkmark & \cellcolor{green!30}\checkmark & \cellcolor{green!30}\checkmark & \cellcolor{green!30}\checkmark \\ 
\hline 
Incoherence & \cellcolor{green!30}\checkmark & \cellcolor{green!30}\checkmark & \cellcolor{green!30}\checkmark & \cellcolor{green!30}\checkmark\\ 
\hline 
Phase Locked & \cellcolor{green!30}\checkmark & \cellcolor{green!30}\checkmark & \cellcolor{green!30}\checkmark & \cellcolor{green!30}\checkmark\\ 
\hline 
Two cluster/sync dots & \cellcolor{green!30}\checkmark  & \cellcolor{green!30}\checkmark & \cellcolor{green!30}\checkmark & \cellcolor{green!30}\checkmark \\ 
\hline 
Multi cluster & \cellcolor{green!30}\checkmark  & \cellcolor{yellow!50}?? & \cellcolor{yellow!50}?? & \cellcolor{yellow!50}?? \\ 
\hline 
Split Pinned & \cellcolor{yellow!50}?? & \cellcolor{green!30}\checkmark & \cellcolor{green!30}\checkmark & \cellcolor{green!30}\checkmark\\ 
\hline 
Chimera & \cellcolor{yellow!50}?? & \cellcolor{green!30}\checkmark & \cellcolor{green!30}\checkmark & \cellcolor{green!30}\checkmark \\ 
\hline 
\end{tabular}
\caption{Summary of states realized in different forced swarmalator models. It is observed that the simplified models with periodic boundary conditions do not restrict the diversity of collective behaviors. Instead, they retain the essential mechanisms that drive swarmalator interactions, showing that the tractable models with periodic boundaries are sufficient to capture and study the dynamics of the forced swarmalator systems. } 
\label{table}
\end{table*}

Before proceeding further, to simplify our analysis, we focus exclusively on the case of resonant forcing, where the driving frequency matches the natural frequency, i.e., $\Omega=\omega$. Then by selecting an appropriate reference frame, we set the natural frequencies $u=\nu=w=\omega=0$ without loss of generality. Additionally, by rescaling time, we fix $J=1$. With these assumptions, our models reduce to systems governed by two parameters, $F$ and $K$, through which a rich array of collective dynamics can be explored.
The external forcing wants to pin the phases of the swarmalators at $\theta_{i}=0$ and the phase coupling tend to minimize (for $K>0$) or maximize (for $K<0$) the phase differences between them. However, the overall dynamics become more complex due to the influence of spatial interactions, which modulate the strength of phase interactions among oscillators.

Table \ref{table} summarizes the possible states that emerge in these forced swarmalator models. Notie that all these models reproduce almost similar dynamics in the presence of external forcing.


\section{Numerics}

Here, we present our numerical investigations for the forced swarmalator models. Unless otherwise specified, simulations are conducted by integrating the equations of motion over a time period of $T=1500$ units using Julia’s adaptive ODE solver with an integration tolerance of $10^{-12}$. For the 2D swarmalator model, a population of $N=200$ swarmalators are initially positioned randomly within a square of length $2$ and their phases are randomly drawn from the interval $[-\pi,\pi]$. For all other models (i.e., the models with periodic boundary conditions), we consider a population of $N=1000$ swarmalators with both their positions and phases being uniformly sampled from $[-\pi,\pi]$. In the following, we will discuss the various collective states that emerge in these forced swarmalator models. 

\subsection{2D swarmalator model}

Numerical results reveal that the model converges to six distinct long-term states, each determined by the initial configurations and specific parameter values, as illustrated in Fig.~\ref{2d_model_states}. Of these six states, only one is stationary, while the remaining five exhibit nonstationary dynamics.  
 
\textit{Pinned} [Fig.~\ref{2d_model_states}(a)]. This is the only stationary state observed in the model. In this state, the swarmalators arrange themselves in a circularly symmetric spatial distribution, with their phases pinned to the driving field, $\theta_{i}=0$. This state occurs for $F>0$ when $K \ge 0$. For $K<0$, the system also settles in this state provided that the external forcing is sufficiently strong. 

\textit{Multiple cluster} [Fig.~\ref{2d_model_states}(b)]. As we move from 
$K=0$ into the $K<0$ region, we observe the first unsteady state for small 
$|K|$ and low forcing strength. In this parameter range, the swarmalators typically form phase wave states when no forcing is applied $(F=0)$. With $F$ turned on, the swarmalators still split into multiple clusters, each characterized by distinct phases. Within each cluster, the swarmalators exhibit small amplitude oscillations in both position and phase. Unlike the splintered phase wave state, these clusters do not consistently organize into a circular configuration and swarmalators occasionally move from one cluster to a neighboring one. The number of clusters that emerge in this state depends on the initial conditions and parameter values.    

\textit{Two cluster} [Fig.~\ref{2d_model_states}(c)]. As $F$ is further increased, these clusters merge into two distinct clusters with their phases being $\theta_1 \approx \frac{\pi}{4}$ and $\theta_2 \approx -\frac{\pi}{4}$. 

\textit{State with two coherent phase group} [Fig.~\ref{2d_model_states}(d)]. 
As $K$ becomes more negative, the swarmalators transition into another unsteady state. Here, they form a roughly circular spatial configuration, within which they experience simultaneous attraction and repulsion. Their phases are predominantly locked around two opposing values, $\theta_1 \approx \frac{\pi}{4}$ and $\theta_2 \approx -\frac{\pi}{4}$. However, a subset of the swarmalators exhibit phase drift away from these two primary values. Thus, while most phases are anchored around $\theta_1$ and $\theta_2$, there is still some phase variability within the system. This state can be observed more clearly through the scatter plot in $(\phi,\theta)$ plane (not shown here), where $\tan(\phi)=y/x$.  

\textit{Phase locked} [Fig.~\ref{2d_model_states}(e)]. In the negative $K$ regime, when the external forcing $F$ is adjusted to a value slightly greater than the phase coupling $K$, most of the swarmalators tend to align their phases with the driving field, $\theta_{i}=0$. However, a subset of the swarmalators experiences phase drift, preventing full phase locking and resulting in a nonstationary phase-locked state.

\textit{Unsteady incoherence} [Fig.~\ref{2d_model_states}(f)]. The swarmalators form an unsteady incoherence state achieving nearly all possible phase and spatial configurations, leading to a disordered pattern in both position and phase. This state is observed in the negative $K$ regime when the external forcing is relatively smaller than the phase coupling $K$.

Now, in order to characterize the collective states, we use the following order parameters, 

\begin{equation}
	\begin{array}{l}
		W_{\pm}= S_{\pm}e^{\mathrm{i}\psi_{\pm}}= \frac{1}{N} \sum\limits_{j=1}^{N} e^{\mathrm{i}(\theta_{j}+\phi_{j})}, \\
		R_{\theta}e^{\mathrm{i}\psi_{\theta}}=\frac{1}{N} \sum\limits_{j=1}^{N} e^{\mathrm{i}\theta_{j}}, \\
		\langle V \rangle = \frac{1}{N} \sum\limits_{j=1}^{N} \sqrt{(\dot{x}_{j}^2+ \dot{y}_{j}^2+ \dot{z}_{j}^2)}.
	\end{array}
\end{equation}
$S_{\pm}$, known as the rainbow order parameters \cite{o2017oscillators}, quantify the correlation between spatial positions and phases. These parameters reach a maximum value of $1$ when space and phase are fully correlated and a minimum of $0$ when they are uncorrelated. The Kuramoto order parameter $R_{\theta}$ measures the degree of phase synchronization, achieving its maximal value under complete phase synchrony. Lastly, $\langle V \rangle$ represents the mean velocity of the swarmalator population, allowing us to distinguish between stationary $(\langle V \rangle=0)$ and non-stationary states $(\langle V \rangle>0)$.

\begin{figure*}[t!]
	\centering
	\includegraphics[width=0.8\linewidth]{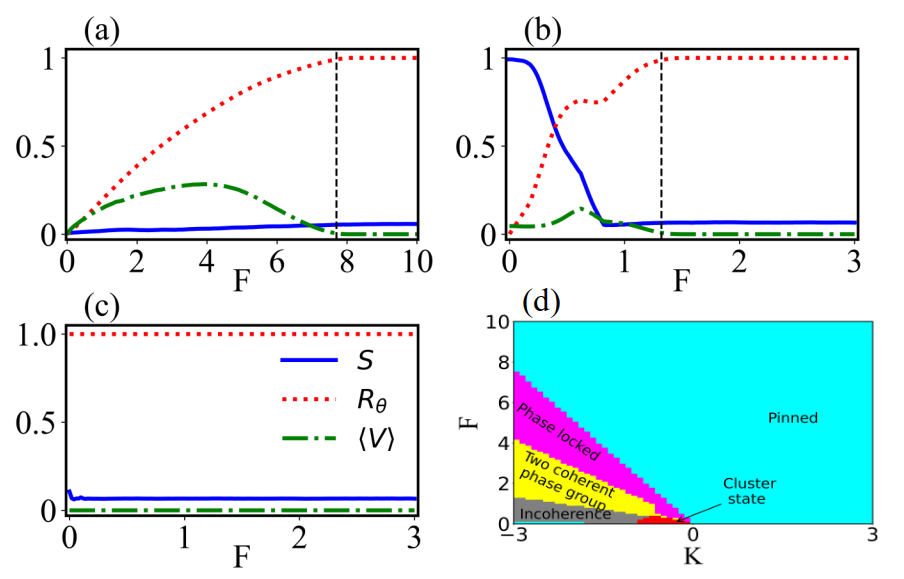}
	\caption{ Variation of order parameters $S=\text{max}\{S_{+},S_{-}\}$, $R_{\theta}$ and mean velocity $\langle V \rangle$ as functions of $F$ for three distinct phase coupling strength in 2D swarmalator model: (a) $K=-3$, (b) $K=-0.5$, and (c) $K=0.5$. The vertical dashed lines in (a) and (b) correspond to the approximate critical forcing strength at which the pinned state emerges. Panel (d) presents a phase diagram in the $(K, F)$ plane for the 2D swarmalator model, where different regions correspond to distinct collective states. Each region is color-coded based on the behavior of the order parameters in the respective states, providing a visual summary of how the model transitions across various collective states as $K$ and $F$ are varied.}
	\label{2d_model_bif}
\end{figure*}

Figure \ref{2d_model_bif} (a-c) illustrates the system’s transitions between different states as $F$ is varied. We categorize the results into three cases based on the values of the phase coupling $K$. These values of 
$K$ are selected to allow the system to display all possible states as $F$ changes.

In Fig. \ref{2d_model_bif}(a), we examine the case of large repulsive phase coupling, $K=-3$. For small $F$, the system exhibits an unsteady incoherent state. Here, the strong repulsive phase coupling drives the swarmalators to maximize phase differences, while the forcing term attempts to align their phases with the driving field. This competition results in unsteady dynamics.
As $F$ increases, the forcing gradually dominates, driving the system to higher speeds, reflected in a monotonic increase in average velocity. However, beyond a critical forcing value $F_{c}$, the mean velocity begins to decrease, while the phase order parameter $R_{\theta}$ rises, signaling the emergence of a state with two coherent phase groups. With further increments in $F$, the mean velocity continues to decline, leading to a phase-locked state that ultimately transitions into a stationary pinned state, where 
$R_{\theta}=1$ and the mean velocity becomes zero.

In Figure \ref{2d_model_bif}(b), we explore the system’s behavior under a smaller repulsive phase coupling, $K=-0.5$. When $F=0$, the spatial and phase distributions of the swarmalators exhibit a strong correlation, yielding phase wave state and rainbow order parameter $S=\text{max}\{S_{+},S_{-}\}$ close to $1$. As $F$ increases, a multi-cluster state begins to form, and 
$S$ declines, indicating a reduction in space-phase correlation. Meanwhile, both the phase order parameter $R_{\theta}$ and mean velocity $\langle V \rangle$ show an upward trend. Upon reaching a critical forcing strength $F_{c}$, the system transitions into a two-cluster state. Here, $S$ approaches a minimum, $R_{\theta}$ stabilizes around $0.7$, and mean velocity starts to decrease.
With further increases in $F$, the two-cluster state shifts into a phase-locked state, where $R_{\theta}$ again rises while mean velocity continues to decrease. At sufficiently high forcing, this phase-locked state ultimately transitions into a pinned state, where $R_{\theta}=1$ and $\langle V \rangle=0$.

For $K>0$, the system consistently reaches the pinned state for any $F>0$. Figure \ref{2d_model_bif}(c) illustrates the behavior of the order parameters as $F$ varies, with $K$ fixed at $0.5$. In this regime, the phases of the swarmalators are effectively pinned to the driving field, resulting in a stationary state characterized by maximal phase synchronization.

Based on the behavior of the order parameters across different states, we outline the approximate regions where each state occurs in the $(K,F)$ parameter plane, as shown in Fig.~\ref{2d_model_bif}(d). However, due to the model’s inherent complexities, analytically deriving stability or existence boundaries for these states remains challenging. Next, we turn our attention to a relatively simplified 2D swarmalator model with periodic boundary conditions. This simplified form allows for greater theoretical tractability, offering insights into the system's behavior under more controlled assumptions. 

\subsection{2D swarmalator model with periodic boundary conditions}

\begin{figure}[t!]
	\centering
	\includegraphics[width=\linewidth]{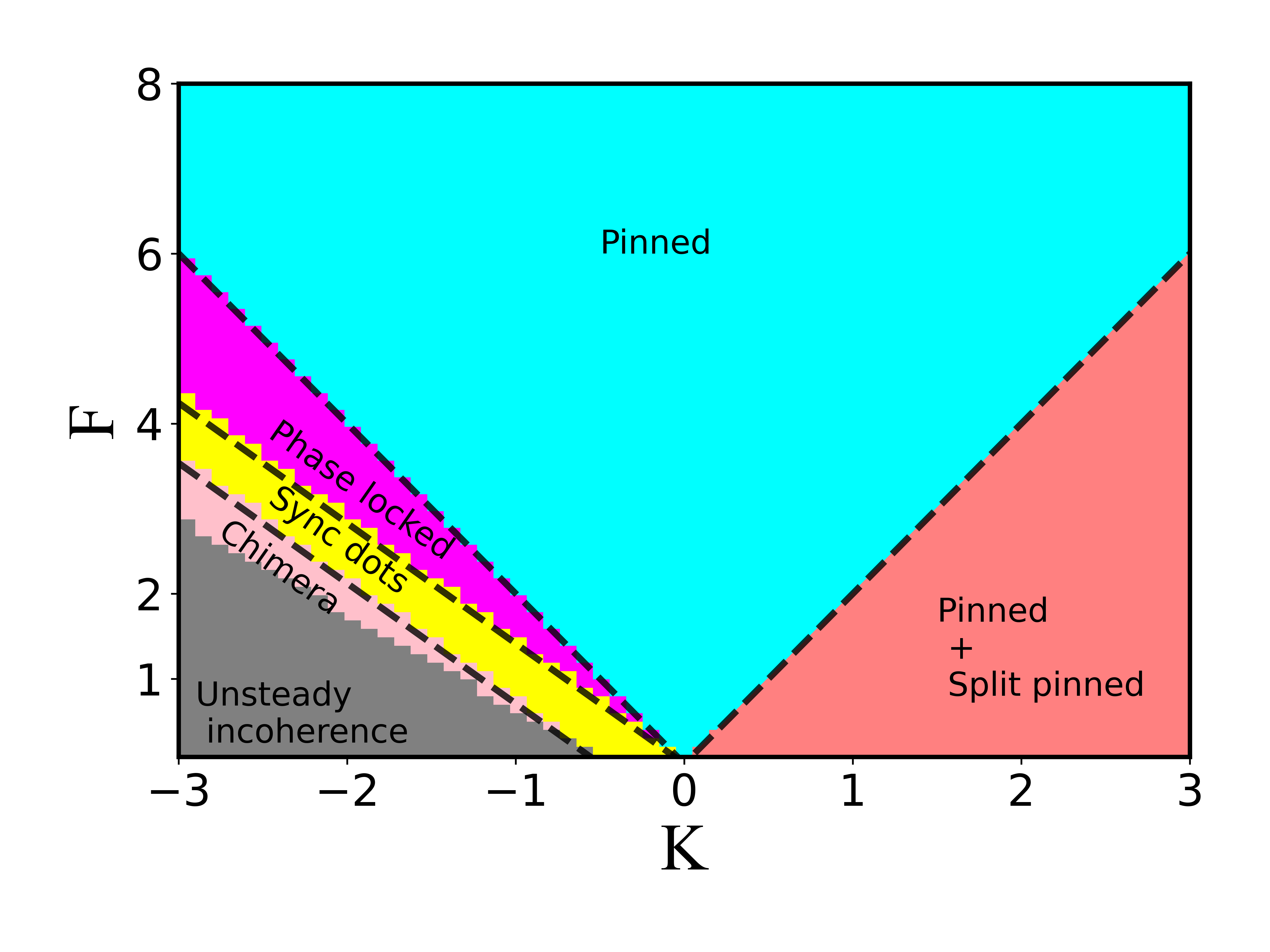}
	\caption{Phase diagram in the $(K,F)$ plane for the 2D swarmalator model with periodic boundary condition. The regions of different states are color-coded according to the behavior of different order parameters in those states by integrating the equation of motions Eq.~\eqref{2d_torus_model_eq}. The dashed black lines are the analytical boundaries of the static states obtained from Eqs. \eqref{2d_pinned_critical}, \eqref{2d_split_pinned_critical}, \eqref{2d_sync_dots_critical} and \eqref{2d_phase_locked_critical}.}
	\label{2d_torus_bif}
\end{figure}

Numerical results indicate that this simplified model, despite lacking the hard-shell repulsion present in the full 2D model, is still able to reproduce similar dynamics. For varying $K$ and $F$, this model exhibits six distinct long-term states, four of which are static and two nonstationary [see Fig.~\ref{2d_torus_states}]. Figure \ref{2d_torus_bif} shows where these states emerge in the $(K,F)$ parameter plane. 

\textit{Pinned} [Fig.~\ref{2d_torus_states}(a), the right most green dot]. As in the 2D model, the phases of the swarmalators in this simplified version are pinned to the driving field, $\theta_{i}=0$. However, due to the absence of hard-shell repulsion, the swarmalators fully aggregate in space, such that $x_{i}=C_{x}$ and $y_{i}=C_{y}$ for some constants $C_{x}$ and $C_{y}$. These constants arise from the rotational symmetry in the spatial equations.

\textit{Split Pinned} [Fig.~\ref{2d_torus_states}(a), blue and left green dots]. The swarmalator population divides into two groups: in one group, phases are pinned at $\theta_{i}=0$, while in the other, they are pinned at 
$\theta_{i}=\pi$. The number of swarmalators in each group varies depending on the initial conditions.

\textit{Sync dots} [Fig.~\ref{2d_torus_states}(b)]. Here, the swarmalators settles into two sync dots with coordinates $(x_{1},y_{1},\theta_{1})=(C_{x},C_{y},\theta^{*})$ and $(x_{2},y_{2},\theta_{2})=(C_{x}+\Delta x,C_{y}+ \Delta y,-\theta^{*})$. The number of swarmalators within each dots varies depending on the initial conditions. The value of $\theta^{*}$ can be found by substituting the fixed point condition as $\theta^{*}=\pi/4$.

\textit{Phase locked} [Fig.~\ref{2d_torus_states}(c)]. In this state, the swarmalators converge to a single spatial position $(x_{i},y_{i})=(C_{x},C_{y})$ but exhibit phase locking, characterized by constant phase differences $\theta_{i}-\theta_{j}=C_{ij}$. This configuration is best visualized through scatter plots in the $(x,\theta)$ and $(y,\theta)$ plane (like the one in Fig.~\ref{1d_model_states}(c)), where the distinct phase differences become evident.

\textit{Chimera} [Fig.~\ref{2d_torus_states}(d)]. In this dynamic state, the swarmalator population divides into three groups: two coherent groups and one incoherent group. The swarmalators in the coherent groups oscillate slowly with small amplitudes, while those in the incoherent group move more rapidly with larger amplitudes, creating a messy region around the coherent clusters. For certain parameter values, the swarmalators within the coherent groups merge into two synchronized clusters, appearing as two distinct sync dots (shown in magenta). We refer to this state as a chimera state, as it generalizes the chimera behavior observed in traditional oscillator systems, where coherent and incoherent dynamics coexist within the same population.

\textit{Unsteady incoherence} [Fig.~\ref{2d_torus_states}(e)]. In this state, the swarmalators form a disordered, nonstationary configuration resembling a gas cloud. Their positions and phases lack coherence, resulting in a structure with no discernible spatial or phase order.

\begin{figure}[t!]
	\centering
	\includegraphics[width=\linewidth]{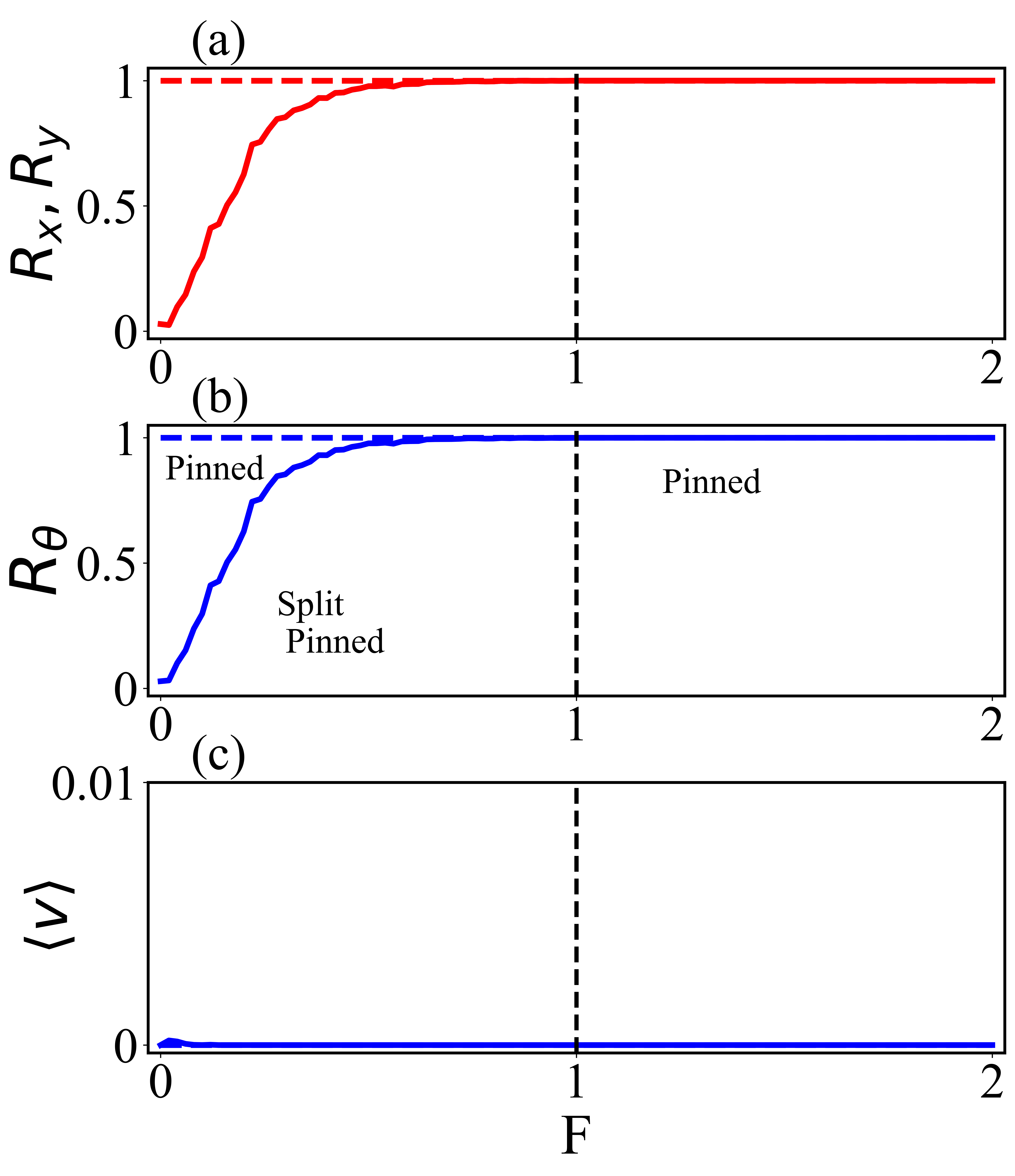}
	\caption{Order parameters $R_{x},R_{y}, R_{\theta}$ and mean velocity $V$ as functions of $F$ for attractive phase coupling $K=0.5$ in the 2D model with periodic boundary conditions. The variations of rainbow order parameters are not depicted as they do not provide any additional information. The dashed vertical line (obtained from Eq.~\eqref{2d_split_pinned_critical}) to the critical coupling below which both pinned and split pinned states exist and beyond which the system only achieves the stable pinned state.}
	\label{2d_torus_ops_vs_F_K_0.5}
\end{figure}

\begin{figure}[t!]
	\centering
	\includegraphics[width=\linewidth]{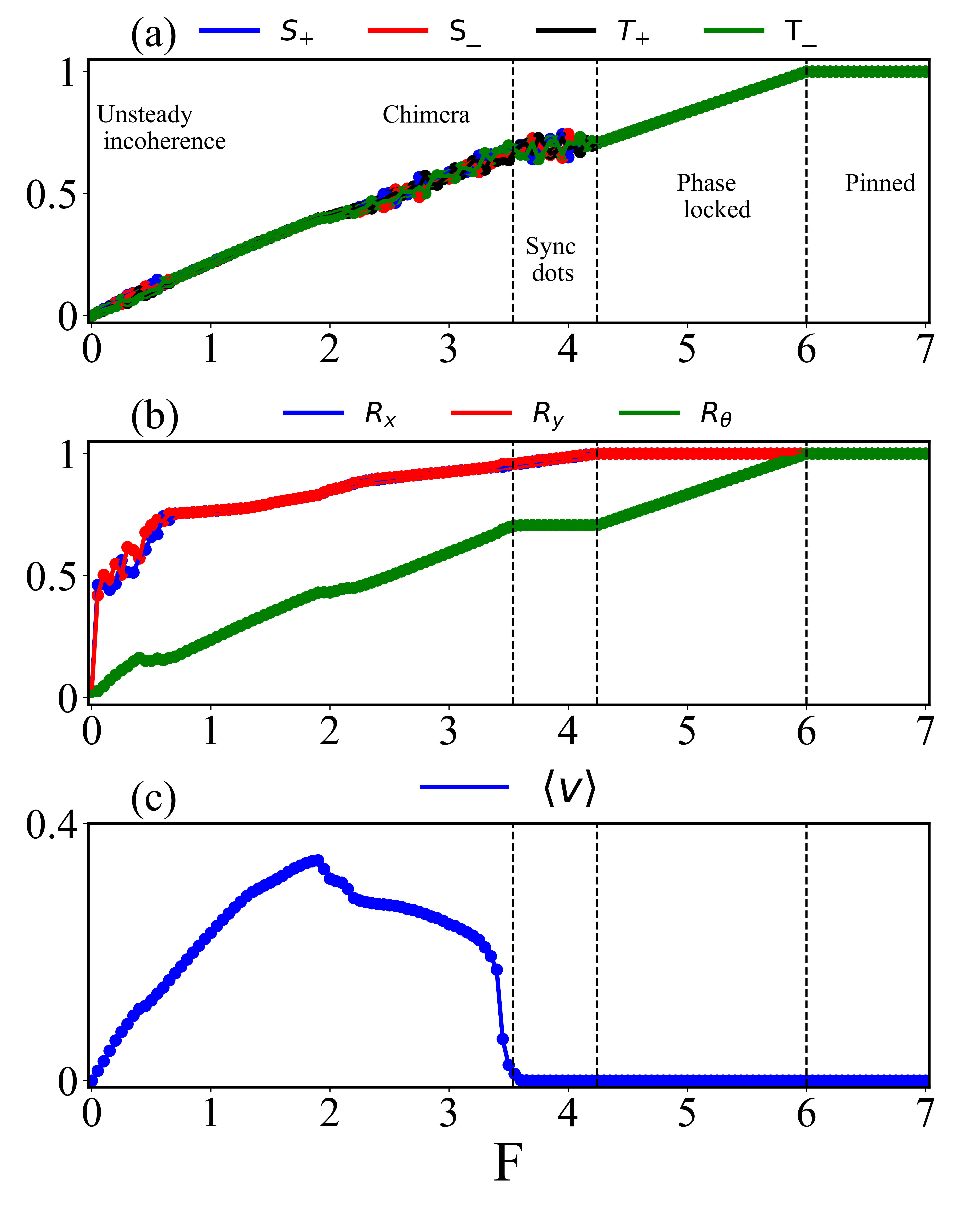}
	\caption{Varaition of different order parameters as functions of forcing strength $F$ for repulsive phase coupling $K=-3$ in the 2D model with periodic boundary conditions. The vertical lines in each panel correspond to the analytically predicted threshold for the three static states (pinned, phase locked, and sync dots), obtained from Eqs.~\eqref{2d_pinned_critical}, \eqref{2d_sync_dots_critical} and \eqref{2d_phase_locked_critical}. }
	\label{2d_torus_ops_vs_F_K_-3}
\end{figure}

In order to characterize the collective states of this model, we use the following order parameters
\begin{equation}
	\begin{array}{l}
		(W_{\pm}, Z_{\pm})= (S_{\pm}e^{\mathrm{i}\psi_{\pm}}, T_{\pm}e^{\mathrm{i}\kappa_{\pm}})= \Big(\dfrac{1}{N} \sum\limits_{j=1}^{N} e^{\mathrm{i}(x_{j}+\theta_{j})},\dfrac{1}{N} \sum\limits_{j=1}^{N} e^{\mathrm{i}(y_{j}+\theta_{j})} \Big), \\
		(R_{x}e^{\mathrm{i}\psi_{x}},R_{y}e^{\mathrm{i}\psi_{y}}, R_{\theta}e^{\mathrm{i}\psi_{\theta}})=\Big(\dfrac{1}{N} \sum\limits_{j=1}^{N} e^{\mathrm{i}x_{j}},\dfrac{1}{N} \sum\limits_{j=1}^{N} e^{\mathrm{i}y_{j}},\dfrac{1}{N} \sum\limits_{j=1}^{N} e^{\mathrm{i}\theta_{j}} \Big), \\
		\langle V \rangle = \dfrac{1}{N} \sum\limits_{j=1}^{N} \sqrt{(\dot{x}_{j}^2+ \dot{y}_{j}^2+ \dot{z}_{j}^2)},
	\end{array}
\end{equation}
where $S_{\pm}$ and $T_{\pm}$ are the usual rainbow order parameters that measure space-phase order, $R_{x/y/\theta}$ quantifies the spatial and phase synchrony and $V$ is the mean velocity.  

Figures \ref{2d_torus_ops_vs_F_K_0.5} and \ref{2d_torus_ops_vs_F_K_-3} illustrate how the system transitions through various states as measured by order parameters.  

For positive phase coupling, the system can reach either the pinned or split-pinned state, depending on the initial configuration, as long as $F<K$ [Fig.~\ref{2d_torus_ops_vs_F_K_0.5}]. These two states are distinguishable by the space and phase coherence order parameters. In the pinned state, all coherence parameters are maximal, i.e., $R_{x}=R_{y}=R_{\theta}=1$. In the split-pinned state, however, $R_{x},R_{y},R_{\theta}<1$ indicating reduced coherence. For $F>K$, only the pinned state remains stable, with all swarmalators aligning in both space and phase.  

For sufficiently large negative phase coupling, the system transitions through all states except the split-pinned state as the external forcing strength $F$ increases [Fig.~\ref{2d_torus_ops_vs_F_K_-3}]. With strong repulsive phase coupling, the swarmalators attempt to maximize phase differences, leading to a uniform distribution in $x,y,\theta$, which results in $S_{\pm}=T_{\pm}\approx 0$. However, the external forcing simultaneously tries to pin the phases to the driving field, creating a tension that produces an unsteady incoherent state. As $F$ increases, the external driving accelerates the system, causing a monotonic rise in both the mean velocity and order parameters. Upon reaching a critical forcing threshold $F_{c}$, the mean velocity begins to decrease, while other order parameters continue to increase, marking the emergence of the chimera state. With further increases in $F$, the chimera state transitions to a sync dots state, where the mean velocity drops to zero and the phase coherence parameter $R_{\theta}$ stabilizes at $1/\sqrt{2}$. As $F$ continues to grow, the sync dots state bifurcates into a phase-locked state, where $S_{\pm}=T_{\pm}=R_{\theta}<1$ while $R_{x}=R_{y}=1$, indicating phase locking with spatial coherence. Finally, for sufficiently large $F$, this phase-locked state transitions into the pinned state, where all coherence parameters reach their maximal values. 

\subsection{3D swarmalator model with periodic boundary conditions}

We now shift focus to a more realistic model where swarmalators move in three-dimensional space. To maintain analytical tractability, we continue to omit the hard-shell repulsion term, similar to the approach used in the 2D model with periodic boundary conditions. The governing equations of motion for this simplified 3D case are given by Eq.~\eqref{3d_torus_model_eq}.

\begin{figure}[t!]
	\centering
	\includegraphics[width=\linewidth]{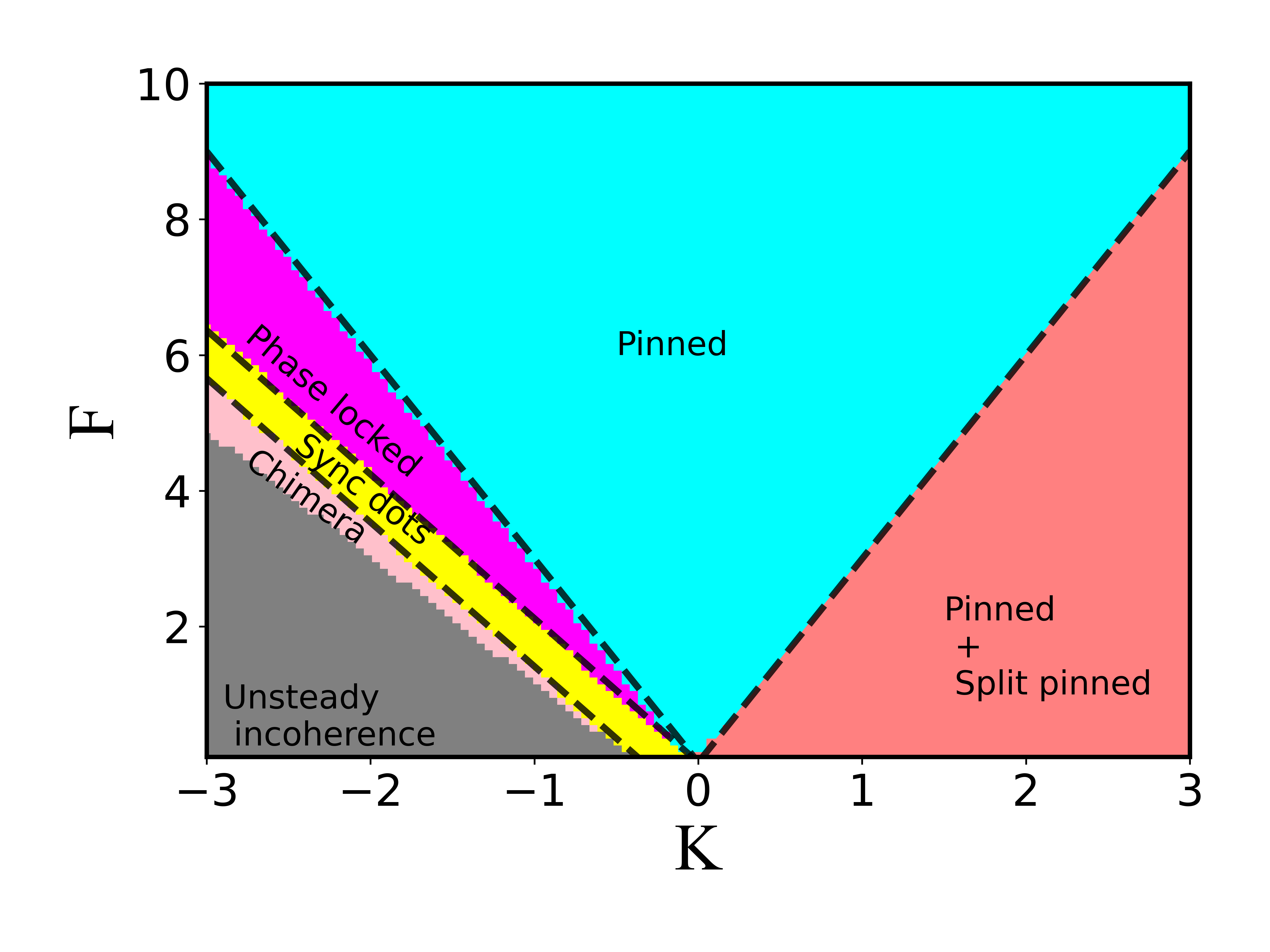}
	\caption{Phase diagram in the $(K,F)$ plane for the 3D swarmalator model with periodic boundary condition. The regions of different states are color-coded according to the behavior of different order parameters in those states by integrating the equation of motions Eq.~\eqref{3d_torus_model_eq}. The dashed black lines are the analytical boundaries of the static states obtained from Eqs. \eqref{3d_pinned_critical}, \eqref{3d_split_pinned_critical}, \eqref{3d_sync_dots_critical} and \eqref{3d_phase_locked_critical}.}
	\label{3d_torus_bif}
\end{figure}

Numerical investigations indicate that this 3D model generates the same long-term collective states as observed in the 2D model with periodic boundary conditions. Figure \ref{3d_torus_bif} maps out the regions in the $(K,F)$ parameter plane where each of these states emerges. Since the scatter plots for the 3D model closely resemble those of the 2D model [Fig.~\ref{2d_torus_states}], we do not show them here. Instead, we will focus on analyzing the collective states that emerge in the 3D model.

\section{Analysis}

In this section, we analyze the static states that emerge in the 2D and 3D models with periodic boundary conditions, deriving the stability boundaries, represented by dashed black lines in the phase diagrams (Figs.~\ref{2d_torus_bif} and \ref{3d_torus_bif}). Given the unsteady nature of the nonstationary states, analytical treatment remains challenging; hence, we are unable to derive stability conditions for these states. We hope future research can provide analytic insights, particularly regarding the swarmalator chimera state, which remains an intriguing and complex behavior to understand.

\subsection{Analysis of the states of the 2D model with periodic boundary conditions}
\subsubsection{Analysis of pinned state}

In the pinned state, the fixed points are
\begin{align}
    x_i &= C_{x},\; y_{i}=C_{y}, \; \theta_i= 0
\end{align}
for some constants $C_{x},C_{y}$. The Jacobian $M_{pinned}$ evaluated at this point has a simple block structure
\begin{equation}
	M_{\text{pinned}} = 
	\begin{bmatrix}
		M_0 & 0   & 0   \\ 
		0   & M_0 & 0   \\
		0   & 0   & M_1 
	\end{bmatrix}, \label{Mpinned}
\end{equation}
where $M_{0}$ and $M_{1}$ are $N \times N$ blocks. Here,
\begin{equation}
	\begin{array}{l}
M_{0_{ij}} = \begin{cases}
	-\dfrac{N-1}{N} & \text{for}\;i=j, \\
	\dfrac{1}{N} & \text{for}\; i \neq j,
\end{cases}
\end{array}  \label{pinned_A0}
\end{equation}
and 
\begin{equation}
	\begin{array}{l}
		M_{1_{ij}} = \begin{cases}
			-K \frac{2(N-1)}{N} - F  & \text{for}\;i=j, \\
			\frac{K}{N} & \text{for}\; i \neq j.
		\end{cases}
	\end{array}  \label{pinned_A1}
\end{equation}
The eigenvalues of $M_{pinned}$ are the sum of the eigenvalues of $M_0$ and $M_1$. $M_0$ has one zero eigenvalue $\lambda_{M_0}$ = 0 stemming from the rotational symmetry of the model and $N-1$ stable eigennvalues $\lambda_{M_0} = -1$. $M_1$ has eigenvalues $-F$ with multiplicity $1$ and $-F-2K$ with multiplicity $N-1$. 

Putting these altogether give the eigenvalues of $M_{pinned}$ as
\begin{equation}
    \begin{array}{l}
     \lambda_0 = 0, \\ 
     \lambda_1 = -1, \\    
     \lambda_2 = -F,  \\
      \lambda_3 = -F-2K,
    \end{array}
\end{equation}
with multiplicities $2$, $(2N-2)$, $1$, $N-1$, respectively. This tells us that the pinned state dies via zero eigenvalue bifurcation at 
\begin{equation}\label{2d_pinned_critical}
    F_c = - 2K,
\end{equation}
which is plotted in the Fig.~\ref{2d_torus_bif}.

\subsubsection{Analysis of split pinned state}
Here, the swarmalators form two clusters with fixed point $(x_1,y_1,\theta_1)=(C_{x},C_{y},0)$ and $(x_2,y_2,\theta_2)=(C_{x}+\pi,C_{y}+\pi, \pi)$. The Jacobian evaluated at the fixed points has a block structure similar to that of the pinned state, i.e.,
\begin{equation}
    M_{spilt \; pinned} = \begin{bmatrix}
    	M_0 & 0   & 0   \\ 
    	0   & M_0 & 0   \\
    	0   & 0   & M_1 
    \end{bmatrix}, \label{MSplitpinned}
\end{equation}
where $M_{0}$ is given as previously by Eq.~\eqref{pinned_A0}, and the elements of the block $M_{1}$ are given by
\begin{equation}
    \begin{array}{l}
         M_{1_{ij}}=\begin{cases}
             \frac{K}{N} & i \neq j \\
             -F-\frac{2(N-1)}{N}K & i=j, \; \text{and} \; 1\leq i \leq [\frac{N}{2}] \\
             F-\frac{2(N-1)}{N}K & i=j, \; \text{and} \; [\frac{N}{2}]+1\leq i \leq N,
            
         \end{cases} 
    \end{array}
\end{equation}
where $[x]$ is the greatest integer less than or equal to $x$. The eigenvalues of $M_{1}$ are 
\begin{equation}
    \begin{array}{l}
     \lambda_1 = -F-2K, \\
     \lambda_2 = F-2K, \\
     \lambda_{3,4} = -K \pm \sqrt{F^2+K^2}, 
    \end{array}
    \end{equation}
with multiplicities $\frac{N}{2}-1$, $\frac{N}{2}-1$ and $1$, respectively. The eigenvalues of $M_{0}$ and $M_{1}$ altogether tells that the split pinned state exists for  
\begin{equation}\label{2d_split_pinned_critical}
    F < 2K \; \text{for}\; K>0.
\end{equation}
The corresponding critical curve is plotted in Fig.~\ref{2d_torus_bif}.
\subsubsection{Analysis of sync dots}
In the sync dots, swarmalators form two groups, and the number of swarmalators in each group depends on the initial conditions. The first group is defined by $(x_1,y_{1}, \theta_1) = (C_{x}, C_{y}. \theta^*)$, while the second is defined by $(x_2,y_{2}, \theta_2)=(C_{x} + \Delta x,C_{y} + \Delta y, -\theta^*)$. The constant $C_{()}$ is arbitrary and stems from the rotational symmetry in the $\dot{x}$ and $\dot{y}$ equations. Here, we analyze the scenario where the swarmalator population splits evenly into two clusters, each containing the same number of swarmalators, without loss of generality. Numerical simulations suggest that when the clusters contain unequal numbers of swarmalators, similar stability properties are observed, indicating that the stability characteristics are likely independent of the precise distribution of swarmalators between the clusters.  The values $(\Delta x, \Delta y, \theta^*)$ can be found by substitution into the equation of motions \eqref{2d_torus_model_eq} as
\begin{equation}
  \begin{array}{l}
       \theta^* = -\pi/4, \\
    \cos \Delta x +\cos \Delta y = -\frac{\sqrt{2} F}{K}.
\end{array}
\end{equation}
To proceed further, we consider any one of the $\Delta x$ or $\Delta y$ to be $0$, which leaves the other as $\cos^{-1}  (-1-\frac{\sqrt{2} F}{K})$ (this is one of the possible fixed point conditions for the sync dots obtained from equation of motions \eqref{2d_torus_model_eq}). Numerics suggests that the acquired stability condition due to the consideration of these fixed points provides the exact stability boundary for the sync dots state.

With all these prerequisites, the eigenvalues of the Jacobian $M_{sync \; dots}$ can be obtained by following the similar procedure as in Ref.~\cite{anwar2024forced}. The eigenvalues are then given by,
\begin{equation}
    \begin{array}{l}
         \lambda_{1}=0, \\
        \lambda_{2}=-\frac{F}{\sqrt{2}}, \\
        \lambda_{3,4}=\frac{-F \sqrt{K} \pm \sqrt{F \left(F (K-16)-16 \sqrt{2} K\right)}}{2 \sqrt{2} \sqrt{K}}, \\
        \lambda_{5,6}= \frac{-K \left(\sqrt{2} F+2 K+1\right) \pm \sqrt{K \left(2 F^2 (K-4)+2 \sqrt{2} F K (2 K-5)+(1-2 K)^2 K\right)}}{4 K}.
    \end{array} 
    \end{equation}
Using this eigenvalues, we can conclude that the state becomes stable when
\begin{eqnarray}\label{2d_sync_dots_critical}
     \Bigg(-K\sqrt{2}-\frac{1}{\sqrt{2}}\Bigg)<F<-K\sqrt{2} \;\; \text{for} \;\;  K<0, 
\end{eqnarray}
which is plotted in Fig.~\ref{2d_torus_bif}.

\subsubsection{Analysis of phase locked state}
The fixed points here take the form
\begin{equation}
\begin{array}{l}
    x_{i}=C_{x}, \\
    y_{i}=C_{y}, \\
    \theta_{i} \in (-a,a),
\end{array}
\end{equation}
where $C_{()}$ and $a$ are constants. Plugging these expressions into the equation of motions \eqref{2d_torus_model_eq} gives us
\begin{eqnarray}
    2KR_{\theta}\sin{(\psi_{\theta}-\theta_{i})}-F\sin{\theta_{i}}=0,
\end{eqnarray}
 Assuming $\psi_{\theta}=0$ without loss of generality, we have 
\begin{equation}
    2KR_{\theta}=-F.
\end{equation}

The state bifurcates from the sync dots configuration, where swarmalators in one group have phases $\theta_{i}=-\frac{\pi}{4}$ and those in the other have $\theta_{i}=\frac{\pi}{4}$. Applying the fixed-point condition for sync dots yields $R_{\theta}=1/\sqrt{2}$. Thus, the phase-locked state bifurcates from the sync dots along the critical curve:
\begin{eqnarray}
	F=-{K}\sqrt{2},
\end{eqnarray}
which aligns with the previously derived stability boundary for the sync dots. The pinned state, defined by $\theta_{i}=0$ (implying $R_{\theta}=1$), bifurcates from the phase locked state along the critical curve 
$F=-2K$, matching its stability boundary.

Therefore, the phase-locked state exists within the range:

\begin{eqnarray}\label{2d_phase_locked_critical}
-K\sqrt{2}<F<-2K \; \text{for}\; K<0.
\end{eqnarray}
This critical boundary is shown in Fig.~\ref{2d_torus_bif}.
This completes our analysis. Figure~\ref{2d_torus_bif} sums up out results in a phase diagram in the $(K,F)$ plane.

\subsection{Analysis of the states of the 3D model with periodic boundary conditions}

\subsubsection{Analysis of pinned state}

In the pinned state, the fixed points are
\begin{align}
	x_i &= C_{x},\; y_{i}=C_{y}, \; z_{i}=C_{z}, \; \theta_i= 0
\end{align}
for some constant $C_{.}$. Proceeding similarly to the analysis of the 2D model, we can obtain the eigenvalues of Jacobian $M_{pinned}$ evaluated at the fixed points as
\begin{equation}
	\begin{array}{l}
		\lambda_0 = 0, \\ 
		\lambda_1 = -1, \\    
		\lambda_2 = -F,  \\
		\lambda_3 = -F-3K,
	\end{array}
\end{equation}
with multiplicities $3$, $(3N-3)$, $1$, $N-1$, respectively. This tells us that the pinned state dies via zero eigenvalue bifurcation at 
\begin{equation}\label{3d_pinned_critical}
	F_c = - 3K,
\end{equation}
which is plotted in Fig.~\ref{3d_torus_bif}.

\subsubsection{Analysis of split pinned state}
Here, the swarmalators form two clusters with fixed point $(x_1,y_1,z_1,\theta_1)=(C_{x},C_{y},C_{z}, 0)$ and $(x_2,y_2,z_2,\theta_2)=(C_{x}+\pi,C_{y}+\pi,C_{z}+\pi, \pi)$. The Jacobian evaluated at the fixed points has the eigenvalues
\begin{equation}
\begin{array}{l}
     \lambda_0 = 0, \\
     \lambda_1 = -1, \\
     \lambda_2 = F-3K, \\
     \lambda_4 = -F-3K, \\
     \lambda_{5,6} = \dfrac{1}{2}\Big(-3K \pm \sqrt{4F^2+9K^2}\Big), 
\end{array}
\end{equation}
with multiplicities $3$, $3N-3$, $\frac{N}{2}-1$, $\frac{N}{2}-1$ and $1$, respectively. The eigenvalues altogether imply that the split pinned state exists for  
\begin{equation}\label{3d_split_pinned_critical}
    F < 3K \; \text{for}\; K>0.
\end{equation}
The corresponding critical curve is plotted in Fig.~\ref{3d_torus_bif}.

\subsubsection{Analysis of sync dots}
Here, as in the 2D system, swarmalators form two groups, and the number of swarmalators in each group depends on the initial conditions. The first group is defined by $(x_1,y_{1},z_{1}, \theta_1) = (C_{x}, C_{y}, C_{z}, \theta^*)$, while the second is defined by $(x_2,y_{2}, z_{2}, \theta_2)=(C_{x} + \Delta x,C_{y} + \Delta y, C_{z} + \Delta z, -\theta^*)$. The values $(\Delta x, \Delta y, \theta^*)$ can be found by substitution into the equation of motions as
\begin{equation}
    \begin{array}{l}
	\theta^* = -\pi/4, \\
	\cos \Delta x +\cos \Delta y + \cos \Delta z = -\frac{\sqrt{2} F}{K}.
    \end{array}
\end{equation}

Considering any two of $\Delta{x},\Delta{y},\Delta{z}$ to be zero and proceeding similarly as the 2D model, we obtain the eigenvalues of the Jacobian $M_{sync \; dots}$ as follows,
\begin{equation}
	\begin{array}{l}
		\lambda_{1}=0, \\
		\lambda_{2}=-\frac{F}{\sqrt{2}}, \\
		\lambda_{3,4}=\frac{-F \sqrt{K} \pm \sqrt{F^2 (K-16)- 32 \sqrt{2} F K-24 K^2}}{2 \sqrt{2} \sqrt{K}}, \\
		\lambda_{5,6}= \frac{-K \left(\sqrt{2} F+3 K+1\right) \pm \sqrt{K \left(2 F^2 (K-4) +6 \sqrt{2} F K (K-3)+9 K^2 (K-2)+K\right)}}{4 K}.
	\end{array} 
\end{equation}
Thus, the state becomes stable when
\begin{eqnarray}\label{3d_sync_dots_critical}
	-\frac{(3K+1)}{\sqrt{2}}<F<-\frac{3K}{\sqrt{2}} \;\; \text{for} \;\;  K<0, 
\end{eqnarray}
which is plotted in Fig.~\ref{3d_torus_bif}.

\begin{table*}[t!]
\centering
\rowcolors{1}{gray!25}{white}
\begin{tabular}{|l|c|c|c|c|} 
\hline 
\rowcolor{gray!50} 
\textbf{State} & \textbf{1D Model (p=1)} & \textbf{2D Model Periodic BC (p=2)} & \textbf{3D Model Periodic BC (p=3)} & \textbf{pD Model Periodic BC ??} \\ 
\hline 
Pinned & $F>-K$ & $F>-2K$ & $F>-3K$ & $F>-pK$ \\ 
\hline 
Split Pinned & $F<K$, $K>0$ & $F<2K$, $K>0$ & $F<3K$, $K>0$ & $F<pK$, $K>0$ \\ 
\hline 
Sync dots & $-\frac{(K+1)}{\sqrt{2}}<F<-\frac{K}{\sqrt{2}}$, $K<0$ & $-\frac{(2K+1)}{\sqrt{2}}<F<-\frac{2K}{\sqrt{2}}$, $K<0$ & $-\frac{(3K+1)}{\sqrt{2}}<F<-\frac{3K}{\sqrt{2}}$, $K<0$ & $-\frac{(pK+1)}{\sqrt{2}}<F<-\frac{pK}{\sqrt{2}}$, $K<0$ \\ 
\hline 
Phase Locked & $-\frac{K}{\sqrt{2}}<F<-K$, $K<0$ & $-\frac{2K}{\sqrt{2}}<F<-2K$, $K<0$ & $-\frac{3K}{\sqrt{2}}<F<-3K$, $K<0$ & $-\frac{pK}{\sqrt{2}}<F<-pK$, $K<0$ \\ 
\hline
\end{tabular}
\caption{Summary of theoretical boundaries of the static states realized in the forced swarmalator models. } 
\label{table2}
\end{table*}


\subsubsection{Analysis of phase locked state}
The fixed points here take the form
\begin{equation}
\begin{array}{l}
	x_{i}=C_{x}, \\
	y_{i}=C_{y}, \\
    z_{i}=C_{z}, \\
	\theta_{i} \in (-a,a),
    \end{array}
\end{equation}
where $C_{()}$ and $a$ are constants. Plugging these expressions into the equation of motions \eqref{3d_torus_model_eq} gives us
\begin{eqnarray}
	3KR_{\theta}\sin{(\psi_{\theta}-\theta_{i})}-F\sin{\theta_{i}}=0.
\end{eqnarray}
Assuming $\psi_{\theta}=0$ without loss of generality, and proceeding similarly to the analysis of the 2D model we obtain that the phase locked state exists when
\begin{eqnarray}\label{3d_phase_locked_critical}
	-\frac{3K}{\sqrt{2}}<F<-3K \; \text{for}\; K<0.
\end{eqnarray}
The corresponding critical curve is delineated in Fig.~\ref{3d_torus_bif}.

This completes our analysis. Table \ref{table2} summarizes the theoretical stability boundaries for the static states observed in the forced swarmalator models.

\section{Discussion}

The main contribution of this study is the development of simplified forced swarmalator models that effectively capture the complex dynamics found in higher-dimensional forced swarmalator systems, while also maintaining analytical tractability. These models display a variety of collective states, and we derive existence boundaries for all states except one. This approach advances us toward the broader objective of solving swarmalator models with power-law kernels, which exhibit more realistic dynamics.

The proposed framework with periodic boundary conditions provides a systematic approach for investigating and analyzing collective states in higher-dimensional models. By adopting a Kuramoto-like phase model with periodic forcing and incorporating spatial-phase coupling, we construct a simplified model based on sines and cosines, which facilitates analytical exploration. For instance, extending the model to a $p$-dimensional spatial framework, as in the 2D and 3D cases, allows us to predict analytical phase boundaries for static states, as summarized in Table \ref{table2}. This framework offers a solid foundation for future studies on swarmalator systems in multiple dimensions.

However, many unexplored avenues for further investigation remain. One key question remains about the analysis of the swarmalator chimera states. We were unable to analyze this state even in the 1D model, where the incoherent subpopulation moves in a peanut-shaped pattern \cite{anwar2024forced}. As spatial dimensions increase here, the motion of the incoherent subset becomes more complex, making analysis more challenging. We hope that future studies could uncover methods to analyze this intriguing state.

In addition, we have focused here on a highly simplified model of forced swarmalators, with identical agents, resonant external forcing, and fixed spatial coupling. To better reflect the complexity of real-world forced swarmalator systems, future work could explore the impact of nonresonant forcing, environmental noise, and nonidentical natural frequencies.


\end{document}